\documentclass[prl,twocolumn,altaffilletter,numerical,flushbottom,secnumarabic,superscriptaddress,floatfix]{revtex4-2}
\usepackage[utf8]{inputenc}
\usepackage[american]{babel}
\usepackage[T1]{fontenc}
\usepackage[pdftex]{graphicx}  
\usepackage{xcolor}
\usepackage{dcolumn}
\usepackage{comment}
\usepackage{bm}
\usepackage{dsfont}
\usepackage{amsthm}
\usepackage{mathrsfs}
\usepackage{bbold}
\usepackage{braket}
\usepackage{wrapfig}
\usepackage{mathtools}
\usepackage{float}
\usepackage{orcidlink}
\usepackage{physics}
\usepackage[normalem]{ulem}
\usepackage{etoolbox}

\usepackage{hyperref}
\hypersetup{colorlinks,bookmarksopen,bookmarksnumbered,citecolor=red,linkcolor=red,urlcolor=red}

\begin{document}

\title{Magic in Grover's Algorithm and the Role of Oracle Synthesis}

\author{Alessio Paviglianiti~\orcidlink{0000-0002-2719-7080}}
\author{Vincenzo Savona}
\affiliation{Institute of Physics, Ecole Polytechnique Fédérale de Lausanne (EPFL), CH-1015 Lausanne, Switzerland}
\affiliation{Center for Quantum Science and Engineering, Ecole Polytechnique Fédérale de Lausanne (EPFL), CH-1015 Lausanne, Switzerland}

\begin{abstract}
Grover's algorithm is a textbook example of quantum advantage, providing a quadratic query speedup over classical methods for unstructured search. In this Letter, we characterize the non-stabilizerness, or \textit{magic}, developed throughout Grover's protocol, quantifying quantum complexity with respect to the efficiently simulable stabilizer states. First, we compute the compilation-independent magic at the end of each Grover iteration, deriving an analytic prediction for the stabilizer Rényi entropy and proving that it remains intensive as the system size grows. We then investigate the magic overhead due to decomposing the circuit into elementary gates. When the solution to the search problem is known, we show that an efficient decomposition exists that keeps magic bounded despite using an extensive number of Toffoli gates. In contrast, agnostic implementations are expected to generate volume-law non-stabilizerness, as we confirm for the $3$-SAT problem, which rules out efficient simulation by stabilizer-based methods in practically relevant cases. Our findings clarify the growth of magic in Grover's algorithm and the role of solution-awareness in its circuit compilation.
\end{abstract}

\maketitle

\textit{Introduction} -- Understanding the boundaries of classical simulation is a central problem in quantum information theory, as it addresses the requirements for genuine quantum advantage and its certification~\cite{bravyi2018,daley2022,eisert2020,hangleiter2023}. Nowadays, this question is commonly assessed within the framework of quantum resource theory~\cite{chitambar2019}. Many-body states can be complex in multiple ways, each quantified by an information-theoretic measure of the corresponding resource. From this perspective, states with low resources are naturally identified with special classes that can be efficiently simulated on classical computers, and thus offer no quantum advantage: the main examples are matrix-product states (MPSs) with low bond dimension~\cite{vidal2003,schollwock2011}, which have bounded entanglement, and stabilizer states~\cite{gottesman1998,aaronson2004}. In recent years, significant effort has been dedicated to understanding how different quantum resources contribute to the overall complexity of quantum states~\cite{liu2022,frau2024,paviglianiti2026}, and to what extent classical simulation can handle their presence~\cite{bravyi2016,qian2024,masotllima2024,fux2025,gu2025}.

Over the past few years, non-stabilizerness~\cite{bravyi2005,veitch2014,howard2017}, also known as \textit{magic}, has gained increasing attention in the study of Hamiltonian and random-circuit models, with applications to quantum phase transitions~\cite{haug2023,lami2023,tarabunga2023}, chaos~\cite{leone2021}, and monitored dynamics~\cite{fux2024,bejan2024,paviglianiti2025,tirrito2025,santos2026}. Stabilizer states, generated by Clifford unitaries, are efficiently simulable, and magic is the resource that must be supplied to produce genuine quantum complexity. From a computational perspective, magic also sets the cost of classical simulation algorithms based on stabilizer expansions~\cite{bravyi2019} and Pauli propagation~\cite{rudolph2026}. Despite its extensive investigation in random circuits, magic in \textit{structured} protocols remains largely unexplored, with only a few studies on Shor's algorithm~\cite{paviglianiti2026_2} and QAOA~\cite{capecci2025,krueger2026,camilleri2026}. In algorithms, gates are arranged in coordinated patterns and can thus generate particular magic signatures, which may depend on the computational task they encode.

In this Letter, we contribute to filling this gap by characterizing magic in Grover's search algorithm~\cite{grover1997,boyer1998,nielsen2010}. Our study is motivated by Ref.~\cite{stoudenmire2024}, which investigates entanglement and discusses the efficiency of MPS simulation: since magic is a distinct resource controlling the cost of different classical simulation methods, it requires a separate analysis. Grover's algorithm iterates an oracle and a diffusion collective operator to eventually generate a superposition of target computational basis states, providing a quadratic query speedup over classical search. We quantify magic through the stabilizer Rényi entropy~\cite{leone2022} (SRE), resolving its growth at three levels of implementation of the circuit. First, we evaluate it at the end of each Grover iteration, providing an analytic prediction for its evolution and showing that it does not scale with the number of qubits. This sets a \textit{compilation-free} baseline, independent of how the collective unitaries are synthesized into elementary gates. We then consider a decomposition into Clifford and Toffoli gates built explicitly from the knowledge of the solution: in this case, magic grows mid-compilation, yet remains bounded despite an extensive number of Toffoli gates, approaching a constant we derive analytically. Lastly, we implement a solution-agnostic oracle for the $3$-SAT problem~\cite{mezard2002}, representing a paradigmatic application of Grover's algorithm, and compile it into the same gate set with reversible logic operations: here, magic exhibits a clear linear scaling with the number of qubits, signaling that the circuit cannot be efficiently simulated with stabilizer-based techniques. Our results show that magic ranges from intensive to extensive based on the information used to synthesize the same oracle unitary, a feature that is not captured by non-Clifford gate counts alone.

\textit{Setup} -- Grover's algorithm addresses the problem of searching a set of $n$-bit bitstrings $\mathcal{S}$. It is implemented by alternating an oracle operator $U_\mathcal{S}$, defined by
\begin{equation}
    U_\mathcal{S}\ket{\mathbf{x}} = \begin{cases}
        -\ket{\mathbf{x}} \qquad \text{if $\mathbf{x}\in \mathcal{S}$},\\
        \ket{\mathbf{x}} \qquad \text{otherwise},
    \end{cases}
\end{equation}
and a diffusion unitary $D = 2\ket{+}\bra{+}-\mathds{1}$, where $\ket{+} = \left(\frac{\ket{0}+\ket{1}}{\sqrt{2}}\right)^{\otimes n}$. In the following, we define $N=2^n$ for conciseness. The product $D U_\mathcal{S}$ acts as a $2$-dimensional rotation in the space spanned by the target $\ket{\psi_\mathcal{S}}=\frac{1}{\sqrt{S}}\sum_{\mathbf{x}\in \mathcal{S}} \ket{\mathbf{x}}$, where $S=|\mathcal{S}|$, and the orthogonal state
\begin{equation}
    \ket{\psi_\perp}=\frac{\ket{+}-\braket{+}{\psi_\mathcal{S}}\ket{\psi_\mathcal{S}}}{\norm{\ket{+}-\braket{+}{\psi_\mathcal{S}}\ket{\psi_\mathcal{S}}}} = \frac{1}{\sqrt{N-S}}\sum_{\mathbf{x}\notin \mathcal{S}}\ket{\mathbf{x}}.
\end{equation}
In particular, defining $\ket{\psi_t} = (D U_\mathcal{S})^t\ket{+}$, we have $\ket{\psi_T}\approx \ket{\psi_\mathcal{S}}$ for $T = \lfloor \pi/4 \sqrt{N/S}\rfloor$.

The family of stabilizer states is generated from $\ket{0}^{\otimes n}$ by acting with Clifford unitaries, generated by the elementary set made by Hadamard, phase, and $\mathrm{CX}$ gates. These states have no magic, and can be simulated efficiently using the stabilizer formalism~\cite{gottesman1998,aaronson2004}. Non-stabilizerness can be injected through non-Clifford unitaries, the standard example being the $T$ gate, and enables the production of more complex states. Throughout this Letter, we will quantify magic using the second SRE
\begin{equation}\label{m2_definition}
    M_2[\ket{\psi}] = -\log_2\left(\sum_\mathrm{P\in\mathcal{P}_n}\frac{\bra{\psi}P\ket{\psi}^4}{N}\right),
\end{equation}
where $\mathcal{P}_n = \{\mathds{1},X,Y,Z\}^{\otimes n}$ is the set of $n$-qubit Pauli strings. The SRE is a monotone of magic~\cite{leone2024}, providing a measure of quantum complexity of the state. In the following, we evaluate the magic of $\ket{\psi_t}$ as a function of $t$, which tracks the growth of non-stabilizerness throughout the algorithm. This quantifies the essential SRE produced by the algorithm \textit{independent of oracle synthesis}, i.e., the intrinsic magic that must be reached by any implementation, regardless of how the oracle is decomposed into elementary gates. We will later also consider the SRE assuming specific compilations of the circuit, showing that in general they generate additional magic and that the synthesis choice plays a crucial role.

\textit{Compilation-independent magic} -- We evaluate how the magic of $\ket{\psi_t}$ evolves with $t$. Numerically, the evolution is implemented by exact state-vector simulations of the protocol, and the SRE is accessed through the recently proposed fast Walsh-Hadamard transform algorithm~\cite{huang2026,sierant2026}. Interestingly, it is possible to derive a closed analytic form for $M_2$, made possible by the special structure of the state $\ket{\psi_t} = \cos(\theta_t)\ket{\psi_\perp} + \sin(\theta_t)\ket{\psi_\mathcal{S}}$, where $\theta_t = (2t+1)\arcsin(\sqrt{S/N})$. In detail, we find the exact formula
\begin{equation}\label{m2_analytic}
    M_2(t) = -\log_2\left(\sum_{k=0}^8 \mu_k \frac{\sin^{8-k}(\theta_t)}{S^{\frac{8-k}{2}}}\frac{\cos^k(\theta_t)}{(N-S)^\frac{k}{2}}\right),
\end{equation}
where $\mu_k$ are constant coefficients that depend on $n$ and on the geometry realized by the target states within the set $\mathcal{S}$ (see End Matter for their explicit expressions). A much simpler approximate form can be obtained by keeping only the two most relevant contributions, yielding $M_2^{(\mathrm{approx})}(t) = -\log_2\left(\cos^8(\theta_t)+2^{-M_2[\ket{\psi_\mathcal{S}}]}\sin^8(\theta_t)\right)$, where $M_2[\ket{\psi_\mathcal{S}}]$ is the SRE of the target state.
\begin{figure}[t!]
  \includegraphics[width=\columnwidth]{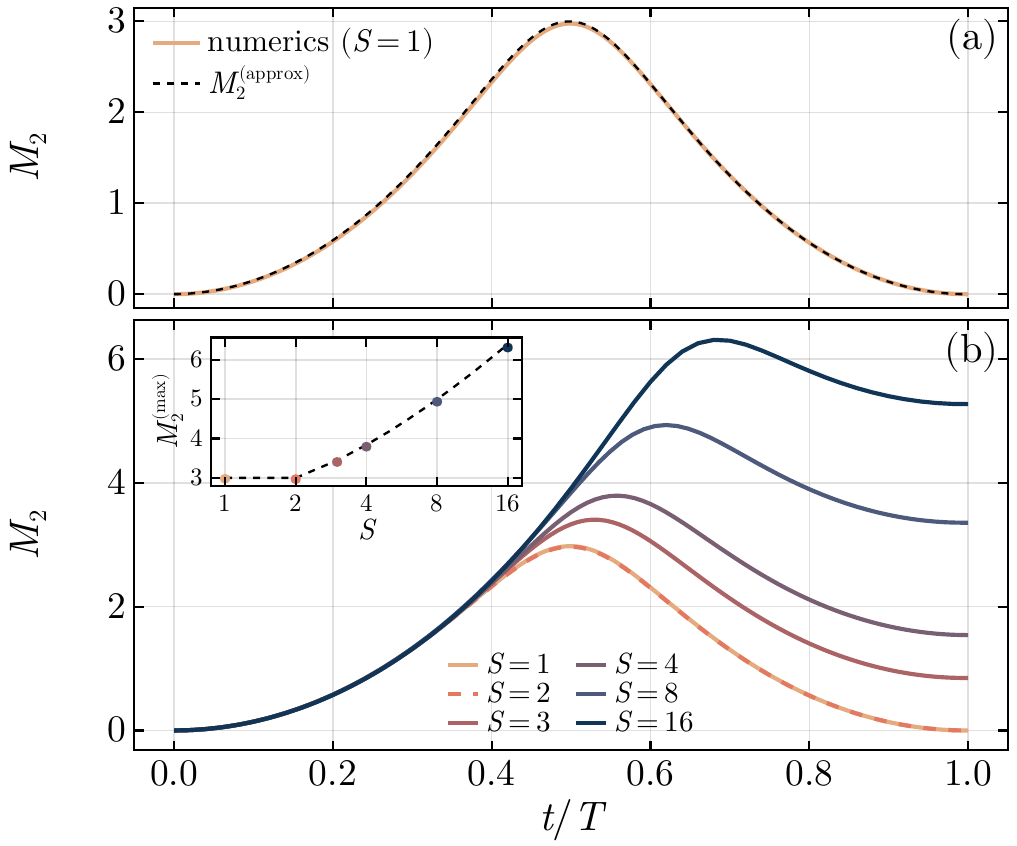}
  \caption{Evolution of magic in Grover's algorithm. (a) Comparison between the analytic approximate formula and exact numerics for $S=1$. (b) Numerical results for multiple choices of $S$, using $\mathcal{S} =\{\ket{3^j \mod N}\}_{j=0,\dots,S-1}$. Inset: maximum of $M_2$ as a function of $S$, compared with the approximate analytic prediction. Data shown is obtained for $n=16$, and numerics yields results indistinguishable from Eq.~\eqref{m2_analytic}.}
  \label{f:essential_magic}
\end{figure}

We present the evolution of magic as a function of $t$ in Fig.~\ref{f:essential_magic}. In particular, Fig.~\ref{f:essential_magic}(a) shows the case of searching a single target state, i.e., $S=1$. The SRE follows a universal profile featuring an initial growth and a late decrease to zero. This is fully expected, as both the initial and final states of the protocol are stabilizers. Notably, the result is independent of the specific bitstring in $\mathcal{S}$, and all choices are equivalent. This also explains why the time evolution is symmetric under time-reversal.

The peak value reached by $M_2$ does not scale with the number of qubits $n$. This observation agrees again with the results derived in previous works for entanglement~\cite{stoudenmire2024,rossi2013,chakraborty2013}, which was found to be intensive as well. These considerations highlight that Grover's algorithm is essentially simple from a classical simulation perspective, as the quantum resources it generates stay bounded as $n$ grows~\footnote{To be precise, neither the entanglement entropy nor the SRE measure the true cost of classical simulation. Still, they lower-bound the corresponding cost metrics for MPS techniques~\cite{vidal2003} and algorithms based on stabilizer expansions~\cite{leone2022}, respectively, and are commonly understood to behave qualitatively similarly as the true cost measures.}. Nevertheless, it is crucial to stress that this does not imply that Grover's algorithm can be simulated efficiently on classical computers for practical applications. Indeed, here the oracle $U_\mathcal{S}$ is implemented as a full $n$-qubit gate, whereas a real implementation would need to synthesize it from elementary reversible logic operations.

When $S>1$ states are searched, the profile of Fig.~\ref{f:essential_magic}(a) is distorted, as shown in Fig.~\ref{f:essential_magic}(b). In particular, the peak value achieved by the SRE grows with $S$, while still remaining intensive as long as $S$ does not scale with $n$. For $S=2$, the target state $\ket{\psi_S}$ is a stabilizer for every $\mathcal{S}$, and the shape of $M_2(t)$ remains unchanged. In contrast, for $S>2$, $\ket{\psi_S}$ can have finite magic depending on the specific bitstrings appearing in the target set.

\textit{Solution-aware oracle decomposition} -- Having understood the synthesis-independent magic baseline of Grover's algorithm, we now move towards the practically relevant case where $U_\mathcal{S}$ and $D$ are decomposed into basic gates. For our purposes, the elementary set we consider consists of Clifford gates, generated by Hadamard, phase, and $\mathrm{CX}$ gates, plus Toffoli gates. For simplicity, we focus on the standard case of $S=1$ throughout the rest of the Letter, and we denote the target bitstring by $\mathbf{x}^*$.
\begin{figure}[t!]
  \includegraphics[width=\columnwidth]{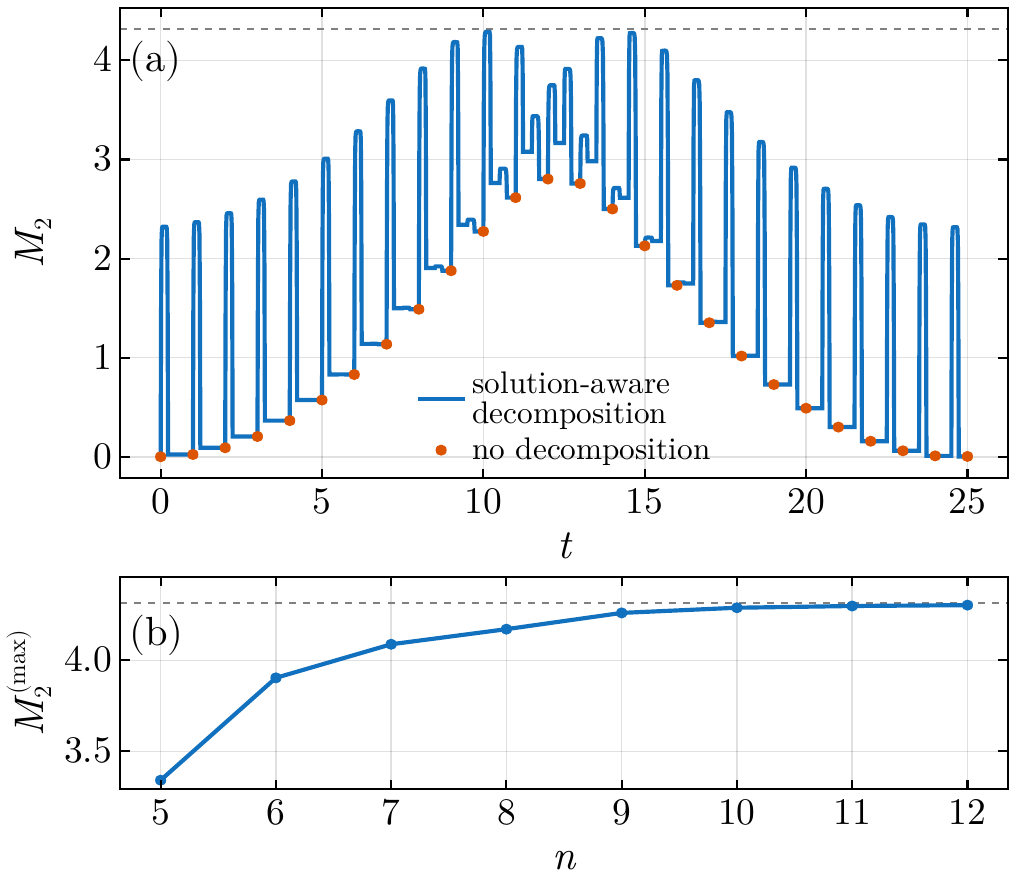}
  \caption{Evolution of magic using the solution-aware decomposition of $\mathrm{C}^{n-1}\mathrm{Z}$ gates. (a) Full simulation for $n=10$, highlighting the compilation-independent values found for integer $t$. (b) Maximum magic achieved at different system sizes. The dashed line marks the asymptotic maximum prediction $3\log_2(1+5^{1/3})$. For reference, the theoretical upper bound to $M_2$ is set by $(n+n_a)$.}
  \label{f:solution_aware_magic}
\end{figure}

Let us assume that the solution $\mathbf{x}^*=(x_1^*,\dots,x_n^*)$ to the search problem is known, in such a way that the oracle can be built leveraging this information. In this case, $U_{\mathrm{x}^*}\equiv U_\mathcal{S}$ admits the relatively simple decomposition
\begin{equation}
    U_{\mathrm{x}^*} = \left(\prod_j X_j^{1-x_j^*}\right) \mathrm{C}^{n-1}\mathrm{Z} \left(\prod_j X_j^{1-x_j^*}\right),
\end{equation}
where $\mathrm{C}^{n-1}\mathrm{Z}$ is the $n$-qubit multi-controlled $Z$ gate. Similarly, the diffusion operator $D$ can be represented as
\begin{equation}
    D = -\left(\prod_j H_j X_j\right) \mathrm{C}^{n-1}\mathrm{Z} \left(\prod_j X_jH_j\right).
\end{equation}
The $\mathrm{C}^{n-1}\mathrm{Z}$ unitary admits a standard decomposition~\cite{nielsen2010} in elementary gates, using $2n-4$ Toffoli gates and $n_a=n-2$ ancilla qubits (indexed by $j=n+1,\dots,n+n_a$). Let $\mathrm{CCX}_{(i_1,i_2),j}$ denote a Toffoli gate with controls $i_1,i_2$ and target $j$. Defining the Toffoli cascade
\begin{equation}\label{toffoli_cascade}
    K^\dagger =\mathrm{CCX}_{(1,2),n+1}\left(\prod_{j=1}^{n-3} \mathrm{CCX}_{(j+2,n+j),n+j+1}\right),
\end{equation}
the multi-controlled $\mathrm{Z}$ gate can be realized as $\mathrm{C}^{n-1}\mathrm{Z} = K^\dagger \mathrm{CZ}_{n,2n-2} K$, with each ancilla initialized in $\ket{0}$.

In the presence of multiple ancilla qubits, the total system size $n_\mathrm{tot}=n+n_a$ grows rapidly, and numerical simulations quickly become infeasible with the standard methods used so far. It is worth noting that the decomposition of $\mathrm{C}^{n-1}\mathrm{Z}$ we implement involves only Toffoli and $\mathrm{CZ}$ gates, thus leaving the total number of computational basis states in the superposition unchanged. As a consequence, it is still possible to store and evolve the state with sparse methods, and magic can be evaluated exactly with the algorithm described in the End Matter.

We compute numerically the evolution of magic throughout Grover's algorithm decomposed in elementary gates, shown in Fig.~\ref{f:solution_aware_magic}(a). In this case as well, the calculation is independent of the specific choice of $\mathbf{x}^*$. As expected, magic can grow significantly throughout the intermediate steps within individual oracle and diffusion calls, thus generating temporary peaks due to the non-Clifford gates of the compilation. Unintuitively, however, Fig.~\ref{f:solution_aware_magic}(b) shows that these peaks do not scale with the number of qubits, thus maintaining magic intensive despite the extensive number of Toffoli gates used at each step. Indeed, the maximum magic approaches $M_2^{(\mathrm{max})} = 3\log_2(1+5^{1/3})$, which can be predicted exactly (see End Matter for the derivation). Hence, the non-Clifford gate count can largely overestimate the magic actually generated.

This result should not be taken to imply that Grover's algorithm remains classically accessible even when accounting for oracle compilation. Indeed, we remark that the oracle decomposition adopted here ``cheats'', in the sense that it leverages the knowledge of the target $\mathbf{x}^*$: as it turns out, this information enables a synthesis route that is sufficiently simple to keep magic low. In practically relevant applications of Grover's algorithm, $\mathbf{x}^*$ is unknown and this specific compilation is thus not possible.

Before proceeding, it is worth mentioning that other ways of decomposing the $\mathrm{C}^{n-1}\mathrm{Z}$ gate exist. For instance, Ref.~\cite{barenco1995} proposes a construction that uses $8(n-4)$ Toffoli gates, but a single ancilla qubit. We implemented this method as well, and we numerically observed that the maximum magic achieved throughout the circuit scales with $n$. This appears to be caused by the recycling of idle qubits to implement Toffoli gates. Indeed, what we showed is that there exists a synthesis strategy that keeps magic intensive, but naturally not all solution-aware compilations produce intensive non-stabilizerness.

\textit{Oracle for the $3$-SAT problem} -- So far, our results indicate that quantum complexity remains bounded when the oracle is either not decomposed or constructed from the solution of the search problem. Now, we finally address the case where $U_{\mathbf{x}^*}$ encodes a search problem whose solution is classically hard to find. In detail, we focus on the 3-SAT problem, for which the oracle must be decomposed through elementary reversible logic operations. As we show below, the resulting magic obeys a volume law.
\begin{figure}[t!]
  \includegraphics[width=\columnwidth]{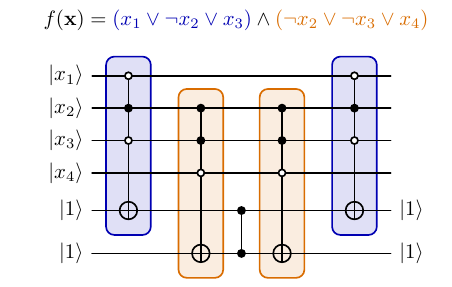}
  \caption{Example of oracle decomposition for the 3-SAT problem. Each clause introduces a pair of $\mathrm{C}^3\mathrm{X}$ gates and an ancilla qubit.}
  \label{f:clause_circuit}
\end{figure}

Given an arbitrary Boolean function $f(\mathbf{x})$, it defines a Grover oracle through $U_\mathcal{S} = \mathrm{diag}\{(-1)^{f(\mathbf{x})}\}$, where $\mathcal{S}$ is the set of inputs for which $f(\mathbf{x})=1$. The 3-SAT problem restricts to a specific class of functions in the form
\begin{equation}\label{3sat_function}
    f(\mathbf{x}) = \prod_{m=1}^{n_c} (\Tilde{x}_{i_m}\lor \Tilde{x}_{j_m}\lor \Tilde{x}_{k_m}),
\end{equation}
where $m$ labels a total of $n_c$ \textit{clauses}, $i_m, j_m, k_m\in\{1,\dots, n\}$ are distinct bit indices, and $\Tilde{x}_i \in \{x_i,\neg x_i\}$. Each clause is \textit{satisfied} (i.e., it returns $1$) if at least one among $\Tilde{x}_{i_m}$, $\Tilde{x}_{j_m}$, and $\Tilde{x}_{k_m}$ is true. The function $f(\mathbf{x})$ then evaluates to $1$ for all bitstrings $\mathbf{x}$ that satisfy all $n_c$ clauses simultaneously. The problem of determining the satisfiability of Eq.~\eqref{3sat_function} belongs to the NP-complete class. In the following, we focus on Grover's oracles induced by 3-SAT functions $f(\mathbf{x})$ admitting exactly one solution $\mathbf{x}^*$, which can be achieved with good probability picking $n_c\approx 4.2 \times n$~\footnote{Given a generic function in the form of Eq.~\eqref{3sat_function}, there is no efficient way of knowing the number of bitstrings that satisfy it a priori. In practice, we draw a random target $\mathbf{x}^*$ and then sample random clauses satisfied by $\mathbf{x}^*$, until $n_c$ are collected. To ensure that a single solution exists, we check it explicitly a posteriori (which is possible at the values of $n$ we consider), and resample the clauses otherwise. For $n\gg 1$, it is conjectured that random $3$-SAT functions become asymptotically unsatisfiable for a density $n_c/n>\rho_c$, where $\rho_c\approx 4.27$~\cite{mezard2002}. Hence, picking a density slightly below this threshold has a high probability of yielding a small number of solutions, at the sizes considered here. The same density was used in Ref.~\cite{stoudenmire2024}.}.

The oracle corresponding to Eq.~\eqref{3sat_function} can be decomposed with a standard compute-uncompute setup, which uses $n_c$ additional ancilla qubits. The circuit, illustrated in Fig.~\ref{f:clause_circuit}, works as follows. A $\mathrm{C}^3\mathrm{X}$ gate using qubits $\{i_m,j_m,k_m\}$ as controls encodes the value of the clause $\Tilde{x}_{i_m}\lor \Tilde{x}_{j_m}\lor \Tilde{x}_{k_m}$ in an ancilla. After encoding all $n_c$ clauses, a $\mathrm{C}^{n_c-1}\mathrm{Z}$ gate is applied to all auxiliary qubits, adding a $\pi$ phase only to those configurations for which all ancillas are in the $\ket{1}$ state. Finally, the ancillas are decoupled from the physical qubits by applying the $\mathrm{C}^3\mathrm{X}$ blocks again. 

\begin{figure}[t!]
  \includegraphics[width=\columnwidth]{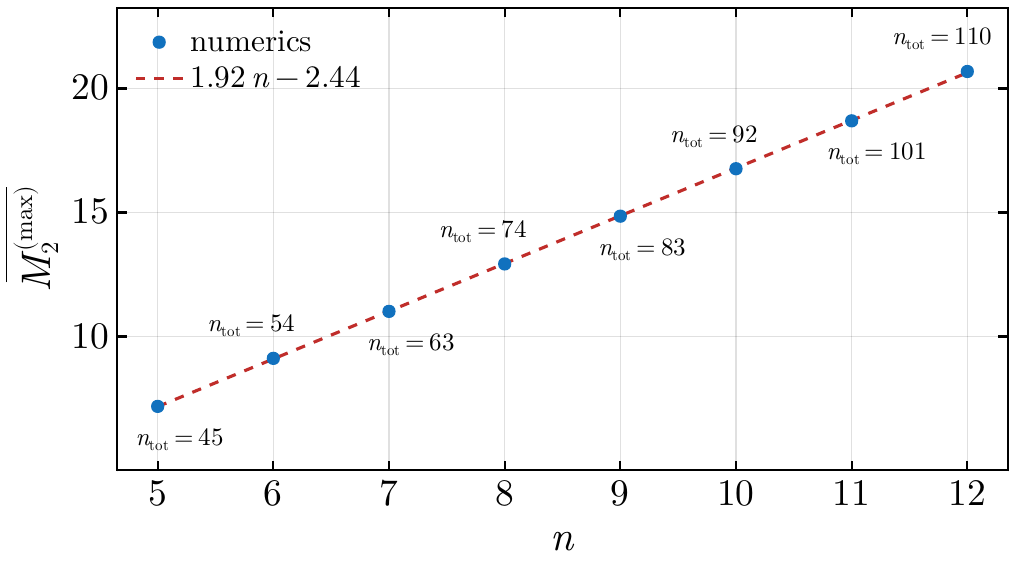}
  \caption{Scaling behavior of the maximal magic achieved throughout the decomposition of the first oracle $U_{\mathbf{x}^*}$, averaged over $10$ random realizations of the 3-SAT problem. Even with these few realizations, the statistical standard errors are smaller than the marker sizes, and are thus omitted. The red line shows the best linear fit. The total qubit number $n_\mathrm{tot}$ for each $n$ is displayed for reference.}
  \label{f:3sat_magic}
\end{figure}

In our implementation, we generate oracles randomly and decompose them as described above. The $\mathrm{C}^3\mathrm{X}$ and $\mathrm{C}^{n_c-1}\mathrm{Z}$ are further decomposed using Toffoli gates through the same method adopted for the solution-aware case: this requires $n_c-2$ additional ancillas, bringing their total number to $n_a=2n_c-2$ and yielding $n_\mathrm{tot}=n+2n_c-2$. We present our numerical results in Fig.~\ref{f:3sat_magic}, showing the highest magic achieved throughout the decomposition of the first oracle layer $U_{\mathbf{x}^*}$ only, averaged over different random realizations. In this case, the circuit develops volume-law non-stabilizerness already at the initial iteration, with an estimated linear behavior $M_2\sim 1.92 n$. This result supports the expectation that, in general, solution-unaware decompositions of the oracle give rise to complex intermediate states, thus preventing the efficient classical simulation of Grover's algorithm for applications of practical relevance. When the solution is unknown, the compilation of the circuit cannot exploit any relevant structure, which directly results in extensive non-stabilizerness.

\textit{Conclusions} -- We have characterized the evolution of magic in Grover's algorithm, providing analytic predictions for the behavior of the SRE. Our findings show that non-stabilizerness is bounded when the oracle and diffusion unitaries are not decomposed, and can always be kept intensive by a suitable solution-aware compilation, despite the need for an extensive number of Toffoli gates. In contrast, agnostic implementations, such as the $3$-SAT oracle, produce large magic scaling linearly with the number of qubits. 

Along the way, adapting the approach of Refs.~\cite{huang2026,sierant2026}, we developed a method for computing the SRE exactly in the presence of ancilla qubits that store classical functions of the computational register, with a cost that does not depend on $n_a$. It can be readily applied to circuit blocks built from reversible logic and diagonal gates acting on arbitrary states, such as the oracles and arithmetic subroutines common in quantum algorithms. This gives access to magic at system sizes that are otherwise prohibitive with conventional methods.

These results are in line with standard expectations on the complexity of the search task. As pointed out by Ref.~\cite{stoudenmire2024}, Grover's algorithm carries little entanglement between iterations, and our analogous observation for magic indicates that any classical hardness must arise from the implementation of the oracle. If a solution-agnostic oracle decomposition generating intensive non-stabilizerness were found for an NP-complete problem, it would be a natural candidate for efficient stabilizer-based simulation: following the discussion of Ref.~\cite{stoudenmire2024}, the solution could then potentially be obtained classically in $\mathrm{poly}(n)$ time, in conflict with the P $\neq$ NP conjecture. In this regard, our results are consistent with the intrinsic complexity of the problem.

Our work opens several directions for future investigation. While we believe that solution-agnostic oracle decompositions for NP-complete problems must generate extensive magic, a formal proof is missing. This is not automatically guaranteed by the belief that P $\neq$ NP: for instance, any sub-linear scaling $M_2\sim n^\alpha$ with $0<\alpha<1$ would remain classically hard for stabilizer-based methods. Another relevant question concerns how circuit synthesis can minimize the resource overhead. As mentioned above, a solution-aware oracle using a single ancilla produced considerably more magic, growing with $n$, than a multi-ancilla implementation: exploring how equivalent decompositions differ in the resources they generate, and how these can be traded for ancilla qubits, is an interesting point. Finally, the impact of noise on magic could be studied at the level of quantum trajectories, following a recent investigation~\cite{menu2026} of entanglement.

\textit{Acknowledgements} -- We thank Emanuele Tirrito and Matteo Seclì for useful discussions. We acknowledge support from the Swiss National Science Foundation through Projects No. 200020\_215172, 200021-227992, and 20QU-1\_215928, and as a part of NCCR SPIN (grant number 225153).

\bibliography{biblio}

\pagebreak
\widetext
\newpage
\begin{center}
\textbf{\large End Matter}
\end{center}

\setcounter{equation}{0}
\setcounter{figure}{0}
\setcounter{section}{0}

\makeatletter
\renewcommand \thesection{S\@arabic\c@section}
\renewcommand{\theequation}{S\arabic{equation}}
\renewcommand{\thefigure}{S\arabic{figure}}

\textbf{Analytic formula for magic} -- We now derive Eq.~\eqref{m2_analytic}. Consider the state $\ket{\psi_t} = \cos(\theta_t)\ket{\psi_\perp}+\sin(\theta_t)\ket{\psi_\mathcal{S}} = \sum_\mathbf{x}\psi_\mathbf{x}\ket{\mathbf{x}}$, where $\psi_\mathbf{x}=\sin(\theta_t)/\sqrt{S}$ if $\mathbf{x}\in\mathcal{S}$ and $\psi_\mathbf{x}=\cos(\theta_t)/\sqrt{N-S}$ otherwise. Introducing the bitstrings $\mathbf{a},\mathbf{b}\in\{0,1\}^n$, we can parameterize Pauli strings in $\mathcal{P}_n$ as $P_{\mathbf{a},\mathbf{b}} = (-i)^{\mathbf{a}\cdot\mathbf{b}} \left(\prod_{j=1}^{n}Z_j^{a_j}\right)\left(\prod_{j=1}^{n}X_j^{b_j}\right)$. Using this representation in the definition of Eq.~\eqref{m2_definition} and performing the sum over $\mathbf{a}$ explicitly, we obtain
\begin{equation}\label{intermediate_formula}
    2^{-M_2(t)} = \sum_{P\in\mathcal{P}_n} \frac{\bra{\psi_t}P\ket{\psi_t}^4}{N} = \sum_{\mathbf{x}_1,\mathbf{x}_2,\mathbf{x}_3,\mathbf{b}\in\{0,1\}^n} \varphi_\mathbf{b}(\mathbf{x}_1)\varphi_\mathbf{b}(\mathbf{x}_2)\varphi_\mathbf{b}(\mathbf{x}_3)\varphi_\mathbf{b}(\mathbf{x}_1\oplus\mathbf{x}_2\oplus\mathbf{x}_3),
\end{equation}
where $\varphi_\mathbf{b}(\mathbf{x}) = \psi^*_\mathbf{x}\psi_{\mathbf{x}\oplus\mathbf{b}}$. Since the wavefunction amplitudes assume only two possible values, Eq.~\eqref{intermediate_formula} immediately leads to Eq.~\eqref{m2_analytic}. Each $\mu_k$ counts the number of ways of choosing the bitstrings $\{\mathbf{x}_1,\mathbf{x}_2,\mathbf{x}_3,\mathbf{b}\}$ such that the octuplet $\{\mathbf{x}_1,\mathbf{x}_2,\mathbf{x}_3,\mathbf{x}_1\oplus\mathbf{x_2}\oplus\mathbf{x}_3,\mathbf{x}_1\oplus\mathbf{b},\mathbf{x}_2\oplus\mathbf{b},\mathbf{x}_3\oplus\mathbf{b},\mathbf{x}_1\oplus\mathbf{x_2}\oplus\mathbf{x}_3\oplus\mathbf{b}\}$ contains exactly $k$ elements of $\{0,1\}^n\setminus\mathcal{S}$ and $8-k$ elements of $\mathcal{S}$. As a consequence, the calculation of $M_2$ reduces to a combinatorial problem. Assuming $S\ll\sqrt{N}$ and discarding all negligible terms in this limit, we find

\begin{minipage}{0.5\linewidth}
\begin{align}
  \mu_0 &= 7S^2-6S+7\Lambda+\Gamma_8 = 2^{-M_2[\ket{\psi_\mathcal{S}}]}S^4, \\ 
  \mu_1 &= 8(\Gamma_7-\Gamma_8),\\
  \mu_2 &= 28 [S(S-1)(S-2)-\Lambda+\Gamma_6-2\Gamma_7+\Gamma_8],\\
  \mu_3 &= 56[\Lambda(S-4)-3\Gamma_6+3\Gamma_7-\Gamma_8],\\
  \mu_4 &= 14N[\Lambda + S(3S-2)],\\
  \nonumber
\end{align}
\end{minipage}
\begin{minipage}{0.5\linewidth}
\begin{align}
  \mu_5 &= 56N[S(S-1)(S-2) - \Lambda],\\
  \mu_6 &= 28N^2S^2,\\
  \mu_7 &= 8N^3S,\\
  \mu_8 &= N^4-\sum_{k=0}^7\mu_k.
\end{align}
\end{minipage}
Some of these coefficients depend on the geometry of the set $\mathcal{S}$, encoded through the irreducible combinatorial terms
\begin{align}
    &\Lambda = \sideset{}{'}\sum_{\mathbf{x}_1,\dots \mathbf{x}_4\in\mathcal{S}}\delta_{\mathbf{x}_1\oplus\mathbf{x}_2,\mathbf{x}_3\oplus\mathbf{x}_4},\label{Lambda}\\
    &\Gamma_6 = \sideset{}{'}\sum_{\mathbf{x}_1,\dots \mathbf{x}_6\in\mathcal{S}}\delta_{\mathbf{x}_1\oplus\mathbf{x}_2,\mathbf{x}_3\oplus\mathbf{x}_4}\delta_{\mathbf{x}_1\oplus\mathbf{x}_2,\mathbf{x}_5\oplus\mathbf{x}_6},\\
    &\Gamma_7 = \sideset{}{'}\sum_{\mathbf{x}_1,\dots \mathbf{x}_7\in\mathcal{S}}\delta_{\mathbf{x}_1\oplus\mathbf{x}_2,\mathbf{x}_3\oplus\mathbf{x}_4}\delta_{\mathbf{x}_1\oplus\mathbf{x}_2,\mathbf{x}_5\oplus\mathbf{x}_6}\delta_{\mathbf{x}_1\oplus\mathbf{x}_3,\mathbf{x}_5\oplus\mathbf{x}_7},\\
    &\Gamma_8 = \sideset{}{'}\sum_{\mathbf{x}_1,\dots \mathbf{x}_8\in\mathcal{S}}\delta_{\mathbf{x}_1\oplus\mathbf{x}_2,\mathbf{x}_3\oplus\mathbf{x}_4}\delta_{\mathbf{x}_1\oplus\mathbf{x}_2,\mathbf{x}_5\oplus\mathbf{x}_6}\delta_{\mathbf{x}_1\oplus\mathbf{x}_3,\mathbf{x}_5\oplus\mathbf{x}_7}\delta_{\mathbf{x}_1\oplus\mathbf{x}_3,\mathbf{x}_6\oplus\mathbf{x}_8},\label{Gamma8}
\end{align}
where the apostrophe denotes the constraint that all bitstrings in the sums are distinct. These quantities have a geometric interpretation. $\Lambda$ counts how many quadruplet configurations of bitstrings form a ``square'', i.e., a $2$-dimensional affine space within $\mathcal{S}$. Similarly, $\Gamma_k$ counts how many $k$-bitstring configurations are arranged as vertices on a ``cube''. These terms capture correlations within the set $\mathcal{S}$, and become negligible if it is generated randomly. Indeed, for a set of $S$ random bitstrings, on average we expect $\Lambda \sim S^4/N$, $\Gamma_6 \sim S^6/N^2$, $\Gamma_7 \sim S^7/N^3$, and $\Gamma_8\sim S^8/N^4$. 

The most important contributions to Eq.~\eqref{m2_analytic} are $\mu_0$ and $\mu_8$, providing the approximate formula $M_2^{(\mathrm{approx})}(t) = -\log_2\left(\cos^8(\theta_t)+2^{-M_2[\ket{\psi_\mathcal{S}}]}\sin^8(\theta_t)\right)$. This simplified form enables the analytic estimate of the peak value of magic, yielding $M_2^{(\mathrm{max})} \approx M_2[\ket{\psi_\mathcal{S}}]+3\log_2\left(1+2^{-M_2[\ket{\psi_\mathcal{S}}]/3}\right)$. If $\mathcal{S}$ is made of uncorrelated bitstrings, the combinatorial terms of Eqs.~\eqref{Lambda}-\eqref{Gamma8} vanish: in this case, $M_2[\ket{\psi_\mathcal{S}}]=2\log_2 S - \log_2(7-6/S)$, and the difference between the maximum and this value goes to zero as $S$ increases.

\textbf{Calculation of magic in the presence of ancilla qubits} -- Consider the state $\ket{\psi} = \sum_\mathbf{x}\psi_\mathbf{x}\ket{\mathbf{x}}\ket{\mathbf{y}(\mathbf{x})}$, where $\mathbf{x}$ and $\mathbf{y}(\mathbf{x})$ are bitstrings of lengths $n$ and $n_a$, respectively. Proceeding as done in the derivation of Eq.~\eqref{intermediate_formula}, we introduce the bitstrings $\mathbf{a},\mathbf{b}\in \{0,1\}^n$ and $\mathbf{p},\mathbf{q}\in \{0,1\}^{n_a}$, referring to physical and ancilla qubits respectively, to parameterize Pauli strings in $\mathcal{P}_{n_\mathrm{tot}}$ as $P_{(\mathbf{a},\mathbf{p}),(\mathbf{b},\mathbf{q})}$. Following the same steps performed above, we then obtain
\begin{equation}\label{explicit_sum}
    \sum_{P\in\mathcal{P}_{n_\mathrm{tot}}} \frac{\bra{\psi}P\ket{\psi}^4}{2^{n_\mathrm{tot}}} = \sum_{\mathbf{b}\in\{0,1\}^n}\sum_{\mathbf{q}\in\{0,1\}^{n_a}}\sum_{\mathbf{x}_1,\mathbf{x}_2,\mathbf{x}_3,\mathbf{x}_4\in\{0,1\}^n}w_{\mathbf{b},\mathbf{q}}(\mathbf{x}_1)w_{\mathbf{b},\mathbf{q}}(\mathbf{x}_2)w_{\mathbf{b},\mathbf{q}}(\mathbf{x}_3)w_{\mathbf{b},\mathbf{q}}(\mathbf{x}_4)\Delta(\mathbf{x}_1,\mathbf{x}_2,\mathbf{x}_3,\mathbf{x}_4),
\end{equation}
where we introduced $\Delta(\mathbf{x}_1,\mathbf{x}_2,\mathbf{x}_3,\mathbf{x}_4) = \delta_{\mathbf{x}_1\oplus\mathbf{x}_2,\mathbf{x}_3\oplus\mathbf{x}_4}\delta_{\mathbf{y}(\mathbf{x}_1)\oplus\mathbf{y}(\mathbf{x}_2),\mathbf{y}(\mathbf{x}_3)\oplus\mathbf{y}(\mathbf{x}_4)}$ and $w_{\mathbf{b},\mathbf{q}}(\mathbf{x}) = \varphi_\mathbf{b}(\mathbf{x})\delta_{\mathbf{q},\mathbf{y}(\mathbf{x})\oplus \mathbf{y}(\mathbf{x}\oplus \mathbf{b})}$ for brevity. Each coefficient $w_{\mathbf{b},\mathbf{q}}(\mathbf{x}_i)$ adds a constraint on the common bitstring $\mathbf{q}$, and it is thus useful to regroup $\{0,1\}^n$ into subsets that can satisfy the four conditions simultaneously. We define $\mathcal{Q}_{\mathbf{b},\mathbf{q}} = \{\mathbf{x}\in\{0,1\}^n \,|\,\mathbf{y}(\mathbf{x})\oplus \mathbf{y}(\mathbf{x}\oplus \mathbf{b}) = \mathbf{q}\}$, allowing us to rewrite Eq.~\eqref{explicit_sum} as
\begin{equation}\label{subset_sum}
\sum_{P\in\mathcal{P}_{n_\mathrm{tot}}} \frac{\bra{\psi}P\ket{\psi}^4}{2^{n_\mathrm{tot}}} = \sum_{\mathbf{b}\in\{0,1\}^n}\sum_{\mathbf{q}\in\{0,1\}^{n_a}}\sum_{\mathbf{x}_1,\mathbf{x}_2,\mathbf{x}_3,\mathbf{x}_4\in\mathcal{Q}_{\mathbf{b},\mathbf{q}}}\varphi_\mathbf{b}(\mathbf{x}_1)\varphi_\mathbf{b}(\mathbf{x}_2)\varphi_\mathbf{b}(\mathbf{x}_3)\varphi_\mathbf{b}(\mathbf{x}_4)\Delta(\mathbf{x}_1,\mathbf{x}_2,\mathbf{x}_3,\mathbf{x}_4).
\end{equation}
Finally, we consider all the possible sums $\mathbf{k}=\big(\mathbf{x}_1\oplus\mathbf{x}_2,\mathbf{y}(\mathbf{x}_1)\oplus\mathbf{y}(\mathbf{x}_2)\big)\in\{0,1\}^{n_\mathrm{tot}}$ that can be achieved by combining $\mathbf{x}_1,\mathbf{x}_2 \in \mathcal{Q}_{\mathbf{b},\mathbf{q}}$, and we define $C_{\mathbf{b},\mathbf{q}}(\mathbf{k})=\sum_{\mathbf{x}_1,\mathbf{x}_2\in\mathcal{Q}_{\mathbf{b},\mathbf{q}}}\varphi_\mathbf{b}(\mathbf{x}_1)\varphi_\mathbf{b}(\mathbf{x}_2)\,
\delta_{\mathbf{k},(\mathbf{x}_1\oplus\mathbf{x}_2,\mathbf{y}(\mathbf{x}_1)\oplus\mathbf{y}(\mathbf{x}_2))}$. It is easily checked by substitution that $\sum_\mathbf{k} C^2_{\mathbf{b},\mathbf{q}}(\mathbf{k})$ reproduces the quadruple sum over the $\mathbf{x}_i$ bitstrings in Eq.~\eqref{subset_sum}, thus leading to
\begin{equation}\label{m2_from_subsets}
    M_2[\ket{\psi}]=-\log_2\left(\sum_{\mathbf{b}\in\{0,1\}^n}\sum_{\mathbf{q}\in\{0,1\}^{n_a}}\sum_{\mathbf{k}\in\{0,1\}^{n_\mathrm{tot}}} C^2_{\mathbf{b},\mathbf{q}}(\mathbf{k})\right).
\end{equation}
For each $\mathbf{b}$, there are at most $2^n$ choices of $\mathbf{q}$ that yield a non-empty $\mathcal{Q}_{\mathbf{b},\mathbf{q}}$ and thus provide a non-zero contribution. For each $\mathcal{Q}_{\mathbf{b},\mathbf{q}}$, the full set of non-vanishing coefficients $C_{\mathbf{b},\mathbf{q}}(\mathbf{k})$ is obtained with $|\mathcal{Q}_{\mathbf{b},\mathbf{q}}|^2$ operations, which is also the maximum number of them. Overall, the computational cost of evaluating Eq.~\eqref{m2_from_subsets} scales as $\mathcal{O}(8^n)$ in the worst case scenario, independent of $n_a$. For generic $\mathbf{y}(\mathbf{x})$ and $n_a\gtrsim n$, however, each non-empty $\mathcal{Q}_{\mathbf{b},\mathbf{q}}$ typically contains only two elements, except for $\mathbf{b}=\mathbf{0}$ for which $|\mathcal{Q}_{\mathbf{0},\mathbf{0}}|=2^n$: this yields a lower effective cost that scales as $\mathcal{O}(4^n)$.

\textbf{Solution-aware oracle decomposition and its maximum magic} -- Here we derive the asymptotic maximum of the SRE obtained by decomposing the $\mathrm{C}^{n-1}\mathrm{Z}$ gates with the method explained below Eq.~\eqref{toffoli_cascade}. Without any loss of generality, we assume $\mathbf{x}^*=(1,\dots,1)$. When the cascade $K$ is applied to $\ket{\psi_t}\ket{0}^{\otimes (n-2)}$ (as part of $U_{\mathbf{x}^*}$), it produces the state
\begin{equation}\label{toffoli_state}
    K(\ket{\psi_t}\ket{0}^{\otimes (n-2)}) = \sin(\theta_t)\ket{\mathbf{x}^*}\ket{1}^{\otimes (n-2)} + \frac{\cos(\theta_t)}{\sqrt{N-1}}\sum_{\mathbf{x}\in\{0,1\}^n\setminus\{\mathbf{x}^*\}}\ket{\mathbf{x}}\ket{x_1 x_2}\ket{x_1 x_2 x_3}\dots\ket{x_1 x_2 \dots x_{n-1}}.
\end{equation}
For the following calculations, it is convenient to introduce an additional ancilla qubit that stores a copy of the first bit $x_1$, which can be achieved using a $\mathrm{CX}$ gate that does not modify the magic. This allows us to identify $\mathbf{y}(\mathbf{x}) = (x_1,x_1x_2,x_1x_2x_3,\dots,x_1x_2\dots x_{n-1})$, and set $n_a=n-1$. Due to this special form of $\mathbf{y}(\mathbf{x})$, which can only be a string of ones followed by a string of zeros, the SRE can be evaluated exactly by computing $C_{\mathbf{b},\mathbf{q}}(\mathbf{k})$ explicitly in Eq.~\eqref{m2_from_subsets}. We assume $n\gg 1$, and keep only leading-order terms in the following calculations. Let $\ell(\mathbf{x})$ denote the position of the first bit $1$ appearing in $\mathbf{x}$. We find the following non-zero contributions:
\begin{enumerate}{}
    \item For any given $\mathbf{b}$, the choice $\mathbf{q}_{\ell(\mathbf{b}),\tau} = (0,\dots,0,1_{\ell(\mathbf{b})},1,\dots,1_\tau,0,\dots,0)$ corresponds to an affine space $\mathcal{Q}_{\mathbf{b},\mathbf{q}_{\ell(\mathbf{b}),\tau}}$ of size $N/2^{\tau}$. This yields $|\mathcal{Q}_{\mathbf{b},\mathbf{q}_{\ell(\mathbf{b}),\tau}}|$ choices of $\mathbf{k}$ for which $C_{\mathbf{b},\mathbf{q}_{\ell(\mathbf{b}),\tau}}(\mathbf{k})=\cos^4(\theta_t)/(2^\tau N)$. Finally, we sum
    \begin{equation}\label{non_zero_q_contribution}
    \sum_{\mathbf{b}}\sum_{\mathbf{q}\neq\mathbf{0}}\sum_{\mathbf{k}} C^2_{\mathbf{b},\mathbf{q}}(\mathbf{k}) = \sum_\ell \frac{N}{2^\ell}\sum_{\tau=\ell}^{n_a}  \frac{8^{-\tau}}{N}\cos^8(\theta_t) \xrightarrow{n_a\to\infty} \frac{8}{105}\cos^8(\theta_t).
    \end{equation}
    \item For $\mathbf{q}=\mathbf{0}$ and $\mathbf{k}=(\mathbf{a},\mathbf{p})$ with $\mathbf{p}\neq\mathbf{0}$, we leverage the symmetry property $C_{\mathbf{b},\mathbf{q}}((\mathbf{a},\mathbf{p}))=C_{\mathbf{a},\mathbf{p}}((\mathbf{b},\mathbf{q}))$. This can be easily checked to be valid when the state amplitudes are real. Using this and noting that half of the relevant bitstrings $\mathbf{k}$ of the previous case have zeros in all ancilla bits, we obtain half the contribution of Eq.~\eqref{non_zero_q_contribution}.
    \item For $\mathbf{q}=\mathbf{0}$ and $\mathbf{k}=(\mathbf{a},\mathbf{0})$, we obtain non-zero contributions from the set $\mathcal{Q}_{\mathbf{b},\mathbf{0}}$, containing all bitstrings with at least a zero before position $\ell(\mathbf{b})$ and having size equal to $N(1-2/2^{\ell(\mathbf{b})})$. $C_{\mathbf{b},\mathbf{0}}((\mathbf{a},\mathbf{0}))$ can be evaluated by counting how many elements $\mathbf{x}\in \mathcal{Q}_{\mathbf{b},\mathbf{0}}$ fulfill the conditions $\mathbf{x}\oplus\mathbf{a}\in \mathcal{Q}_{\mathbf{b},\mathbf{0}}$ and $\mathbf{y}(\mathbf{x})=\mathbf{y}(\mathbf{x}\oplus\mathbf{a})$, yielding $C_{\mathbf{b},\mathbf{0}}((\mathbf{a},\mathbf{0})) = \cos^4(\theta_t)(1-2/2^{\ell_\text{min}})/N$, where $\ell_\text{min} = \min\{\ell(\mathbf{a}),\ell(\mathbf{b})\}$. Accounting for the trivial contribution from $\mathbf{x}^*$, which yields a macroscopic term $\sin^8(\theta_t)$, the final sum reads
    \begin{equation}
        \sum_{\mathbf{a},\mathbf{b}\in\{0,1\}^n} C^2_{\mathbf{b},\mathbf{0}}((\mathbf{a},\mathbf{0})) = \sin^8(\theta_t)+\sum_{\ell=1}^{n_a} \sum_{\tau=1}^{n_a} 2^{-\ell-\tau}\left(1-\frac{2}{2^{\min\{\ell,\tau\}}}\right)^2\cos^8(\theta_t) \xrightarrow{n_a\to\infty} \sin^8(\theta_t)+\frac{3}{35}\cos^8(\theta_t).
    \end{equation}
\end{enumerate}
Putting all together, we finally obtain $M_2 \approx -\log_2(\sin^8(\theta_t)+\cos^8(\theta_t)/5)$. This formula achieves an intensive peak value of $3\log_2(1+5^{1/3})$ at $\sin^2(\theta_t)=(1+5^{1/3})^{-1}$. Finally, while here we considered the magic in the middle of the oracle decomposition, the same maximum is found for the diffusion unitary, due to the symmetric role of initial and final states in Grover's algorithm.

\end{document}